# Emergent interlayer magnetic order via strain-induced orthorhombic distortion in the 5$d$ Mott insulator Sr$_2$IrO$_4$

S. Shrestha[1], M. Krautloher[2], M. Zhu[3], J. Kim[1], J. Hwang[3], J. Kim[4], J.-W. Kim[4], B. Keimer[2], and A. Seo[1]

[1]Department of Physics and Astronomy, University of Kentucky, Lexington, KY 40506, USA
[2]Max-Planck-Institut für Festkörperforschung, D-70569 Stuttgart, Germany
[3]Department of Materials Science and Engineering, The Ohio State University, Columbus, Ohio 43210, USA
[4]Advanced Photon Source, Argonne National Laboratory, Argonne, Illinois 60439, USA

**Abstract**

We report a La$_2$CuO$_4$-like interlayer antiferromagnetic order in Sr$_2$IrO$_4$ films with large orthorhombic distortion (> 1.5%). The biaxial lattice strain in epitaxial heterostructures of Sr$_2$IrO$_4$/Ca$_3$Ru$_2$O$_7$ lowers the crystal symmetry of Sr$_2$IrO$_4$ from tetragonal ($C_4$) to orthorhombic ($C_2$), guiding the Ir 5$d$ $J_{\text{eff}}$ = 1/2 pseudospin moment parallel to the elongated $b$-axis via magnetic anisotropy. From resonant X-ray scattering experiments, we observed an antiferromagnetic order in the orthorhombic Sr$_2$IrO$_4$ film whose interlayer stacking pattern is inverted from that of the tetragonal Sr$_2$IrO$_4$ crystal. This interlayer stacking is similar to that of the orthorhombic La$_2$CuO$_4$, implying that the asymmetric interlayer exchange interaction along $a$ and $b$-directions exceeds the anisotropic interlayer pseudo-dipolar interaction. Our result suggests that strain-induced distortion can provide a delicate knob for tuning the long-range magnetic order in quasi-two-dimensional systems by evoking the competition between the interlayer exchange coupling and the pseudo-dipolar interaction.

Transition metal oxides are a featured condensed matter system for realizing emergent electronic and magnetic phases due to competing interactions and strong electron correlation. Layered oxides with the quasi-two-dimensional K$_2$NiF$_4$-type structure are of particular interest since they include superconducting high-$T_c$ cuprates [1,2], spin/charge/orbital ordered manganates [3], spin-orbit coupled iridates [4,5], and so forth. Despite their diverse physical properties, the spin Hamiltonian of the system is commonly described by

$$H_{ij} = J_{ij}\, \mathbf{S}_i \cdot \mathbf{S}_j + \mathbf{D}_{ij} \cdot \mathbf{S}_i \times \mathbf{S}_j + \mathbf{S}_i \cdot \Gamma_{ij} \cdot \mathbf{S}_j \qquad \text{Eq.1}$$

, where the first term is the Heisenberg exchange interaction, the second term is the Dzyaloshinskii-Moriya (DM) interaction, and the third term is the pseudo-dipolar interaction, respectively, between two sites $i$ and $j$ [6]. Because of the quasi-two-dimensional nature of the layered oxides, only in-plane interactions are often considered important in theoretical models [7]. This approximation is justified because the Heisenberg exchange interaction within the basal plane is a few orders of magnitude stronger than any interlayer interactions. Nevertheless, it is noteworthy that many layered-oxides exhibit long-range magnetic order with well-defined Néel temperature ($T_N$), which is not allowed in an ideal two-dimensional (2D) Heisenberg spin system with continuous symmetry according to the Mermin-Wagner theorem [8]. The interlayer-long-range magnetic order is determined by interlayer Hamiltonian defined as:

$$H_{\langle i,j\rangle /\!/a} = J_{out}^{a}\, (\mathbf{S}_i \cdot \mathbf{S}_j) + \Gamma_{out}^{a}\, (S_i^a S_j^a - S_i^b S_j^b)$$

$$H_{\langle i,j\rangle /\!/b} = J_{out}^{b}\, (\mathbf{S}_i \cdot \mathbf{S}_j) + \Gamma_{out}^{b}\, (S_i^b S_j^b - S_i^a S_j^a) \qquad \text{Eq.2}$$

, where $J_{out}^{a}$ and $J_{out}^{b}$ are the nearest interlayer exchange interactions and $\Gamma_{out}^{a}$ and $\Gamma_{out}^{b}$ are the anisotropic interlayer pseudo-dipolar interactions along the crystallographic $a$- and $b$-directions, respectively, as shown in Fig 1(a). We define $j_c$ and $\delta_c$ as the energies from the nearest interlayer exchange interaction and the anisotropic interlayer pseudo-dipolar interaction, respectively:

$$j_c \equiv J_{out}^{a}\, (\mathbf{S}_i \cdot \mathbf{S}_j) + J_{out}^{b}\, (\mathbf{S}_i \cdot \mathbf{S}_j)$$

$$\delta_c \equiv \Gamma_{out}^{a}\, (S_i^a S_j^a - S_i^b S_j^b) + \Gamma_{out}^{b}\, (S_i^b S_j^b - S_i^a S_j^a). \qquad \text{Eq.3}$$

Two distinct antiferromagnetic (AFM) interlayer stacking patterns are found in the K$_2$NiF$_4$-type oxides depending on $j_c$ and $\delta_c$. For example, tetragonal La$_2$NiO$_4$ and Sr$_2$IrO$_4$ crystals show the AFM-1 type [9-12] whereas orthorhombic La$_2$CuO$_4$ crystals exhibit the AFM-2 type [13-16], as shown in Fig. 1(b). La$_2$CoO$_4$ crystals present both types depending upon their phase transitions between tetragonal and orthorhombic structures [17].

One may conjecture that the AFM stacking order is correlated with the crystal structure since the AFM-1 (i.e., $\delta_c < 0$) and AFM-2 (i.e., $\delta_c > 0$) type patterns seem to appear in tetragonal and orthorhombic structures, respectively. One way to check is to see if the AFM-2 type order is induced when orthorhombic distortion is applied in a tetragonal system. However, recent experiments by Kim et al. showed that the AFM-1 type order remained in the Sr$_2$IrO$_4$ crystals even under orthorhombic distortion using a piezoelectric strain device [18]. Hence, the interlayer AFM stacking order would be irrelevant to the crystal structure or the magnitude of the orthorhombic distortion (~0.03%) was not large enough to stabilize the AFM-2 type stacking pattern in the Sr$_2$IrO$_4$ crystals. It is formidable to figure out which one is correct because conventional strain/stress device approaches using bulk crystals cannot exert much larger external pressure than the magnitude used in the studies of Ref. [18].

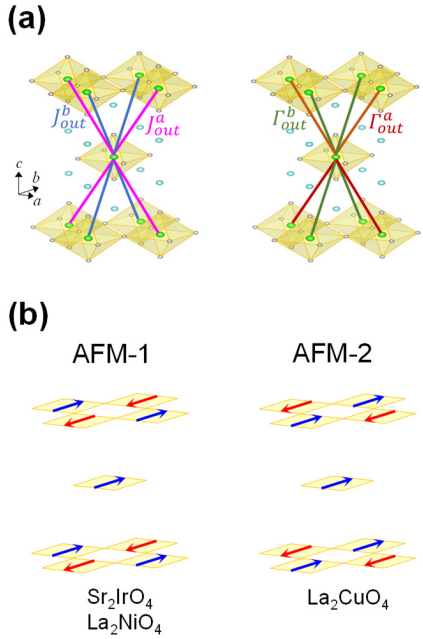

**Fig 1. (a)** Crystal structure of layered oxides with the distorted $K_2NiF_4$ structure. The interlayer exchange interactions $J_{out}^a$ and $J_{out}^b$ share the same interaction paths with interlayer pseudo-dipolar interactions $\Gamma_{out}^b$ and $\Gamma_{out}^b$, respectively. **(b)** Schematic diagrams of two different long-range antiferromagnetic stacking patterns in $K_2NiF_4$ type oxides. AFM-1 type is observed in $Sr_2IrO_4$ and $La_2NiO_4$ whereas AFM-2 is observed in $La_2CuO_4$.

In this *Letter*, we report that the long-range AFM-2 stacking order emerges in the layered iridate $Sr_2IrO_4$ thin films when substantially large (> 1.5%) orthorhombic distortion is applied by biaxial strain. Such a large magnitude of orthorhombic distortion is realized by constructing epitaxial thin-film heterostructures of $Sr_2IrO_4/Ca_3Ru_2O_7$ with excellent structural coherence and atomically sharp interfaces. We observed clear magnetic Bragg reflections from the Ir L-edge resonant X-ray scattering (RXS) experiments, indicating that the AFM-2 type stacking order is stabilized when $|j_c|$ exceeds $|\delta_c|$. Our result confirms that the long-range AFM stacking order of the quasi-two-dimensional oxides is correlated with their crystal structures. It is noteworthy that our experimental finding provides a useful approach for enhancing the magnetic frustrations (i.e., magnetic quantum fluctuations) of various two-dimensional magnetic systems via utilizing the subtle competition between the interlayer exchange interaction and the pseudo-dipolar interaction.

We conceived $Sr_2IrO_4/Ca_3Ru_2O_7$ heterostructures to study the effect of orthorhombic distortion to the AFM order because of the large anisotropic lattice mismatch between the tetragonal $Sr_2IrO_4$ and the orthorhombic $Ca_3Ru_2O_7$, i.e., -2.22% and +0.45% along the *a*- and *b*-directions, respectively, as shown in Fig. 2(a). The heterostructure samples were synthesized by depositing 24-nm-thick $Sr_2IrO_4$ epitaxial thin-films on $Ca_3Ru_2O_7$ (001) single crystals with cleaved surfaces using pulsed laser deposition (PLD). The PLD conditions are a laser fluence of 1.2 J/cm$^2$, a substrate temperature of 700°C, and an oxygen partial pressure of 10 mTorr, which are consistent with Ref. [19]. Figure 2(b) shows X-ray diffraction with well-defined (0 0 $l$) peaks and Kiessig fringes from a Sr$_2$IrO$_4$/Ca$_3$Ru$_2$O$_7$ heterostructure, indicating a high-quality thin film with a thickness of ~24 nm. The structural coherence length of ~29 nm, obtained from the FWHM of the (0 0 20) peak is comparable to the thickness of the Sr$_2$IrO$_4$ thin film (Fig. S1(a)). The H and K scans near (2 0 20) and (0 2 20)-reflections show that the Sr$_2$IrO$_4$ thin films are fully strained along both in-plane directions (Figs. S1(b) and (c)). Large in-plane anisotropy with −2.22% compressive strain along the $a$-axis and +0.45% tensile strain along the $b$-axis is observed in these thin films (Table I) which correspond to the significantly large orthorhombic distortion ($\alpha \equiv (b-a)/(b+a)$) of +1.58% at low temperature (6 K). Considering Young's modulus of ~130 GPa of Sr$_2$IrO$_4$ crystals [20], this compressive strain corresponds to an effective uniaxial pressure of ~2.9 GPa along the $a$-axis, which is far above the accessible range of conventional strain/stress approaches [21]. This anisotropic strain is expected to lower the crystal symmetry of Sr$_2$IrO$_4$ from tetragonal to orthorhombic, as shown in Fig. 2(a). Structural Bragg peaks of (1 2 19) and (1 2 21) that are relevant to the octahedral rotation along the $c$-axis [22,23] were not observed from our samples (Fig. S1(d)), implying that the octahedral rotational pattern is substantially modified in these Sr$_2$IrO$_4$/Ca$_3$Ru$_2$O$_7$ heterostructures presumably due to the large orthorhombic distortion. Figure 2(c) shows high-resolution Z-contrast scanning transmission electron microscopy (STEM) data of the cross-sections of the heterostructures. Note that an atomically sharp interface between Sr$_2$IrO$_4$ and Ca$_3$Ru$_2$O$_7$ is observed without any noticeable misfit dislocations or atomic inter-mixing even under such a large orthorhombic distortion.

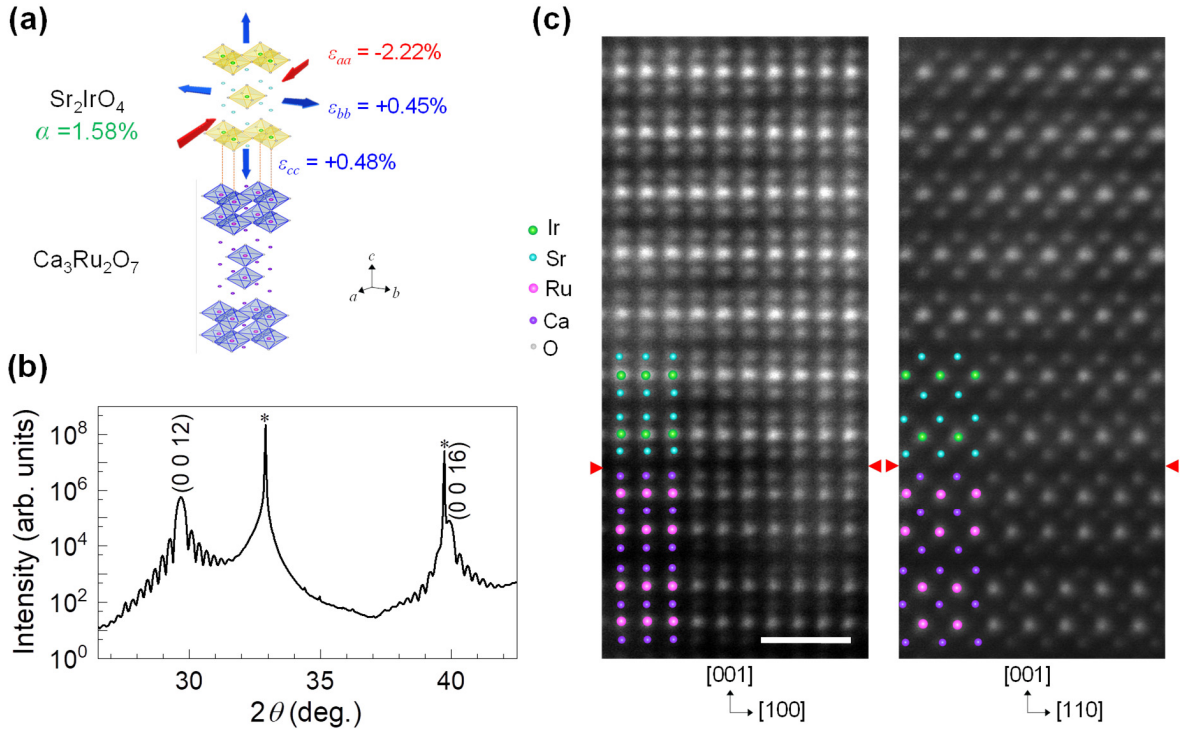

**Fig 2. (a)** A schematic diagram of a $Sr_2IrO_4/Ca_3Ru_2O_7$ heterostructure. The red and blue arrows represent compressive and tensile strain directions and magnitudes. **(b)** X-ray diffraction (0 0 L) scan of a $Sr_2IrO_4/Ca_3Ru_2O_7$ heterostructure. The (0 0 12) and (0 0 18) peaks from the $Sr_2IrO_4$ thin film are visible. The asterisks [∗] indicate the peaks from the $Ca_3Ru_2O_7$ single crystal substrate. **(c)** High-resolution Z-contrast STEM images of a $Sr_2IrO_4/Ca_3Ru_2O_7$ heterostructure for two different cross-sectional directions. The red triangles mark the atomically sharp interface between $Sr_2IrO_4$ and $Ca_3Ru_2O_7$. The scale bar is 1 nm.

Table I: Lattice constants and strain values of Sr$_2$IrO$_4$ (001) thin films grown epitaxially on Ca$_3$Ru$_2$O$_7$ (001) single crystals.

| Epitaxial Sr$_2$IrO$_4$ thin films on Ca$_3$Ru$_2$O$_7$ | $a$ (Å) | $b$ (Å) | $c$ (Å) | α(%)[a] | $\varepsilon_{aa}$ (%) | $\varepsilon_{bb}$ (%) | $\varepsilon_{cc}$ (%) | V (Å$^3$) |
|---|---|---|---|---|---|---|---|---|
| 6 K | 5.364(4) | 5.536(4) | 25.92(2) | **+1.58** | −2.21(7) | +0.93(7) | +0.47(8) | 769.9 |
| 296 K | 5.378(4) | 5.523(4) | 25.94(2) | **+1.33** | −2.22(9) | +0.44(8) | +0.61(8) | 770.5 |

[a] Values are obtained using α ≡ {($b$ − $a$) / ($b$ + $a$)} × 100(%)

Resonant X-ray scattering experiments reveal an unprecedented AFM-2 stacking order, which has not been observed from any Sr$_2$IrO$_4$ crystals or thin films. Using resonant X-ray near the Ir L$_3$ edge ($\hbar\omega$ = 11.217 keV), we observed a clear (1 0 24) magnetic Bragg peak at 6 K with a magnetic coherence length of ~ 26 nm along the out-of-plane direction (Fig. S3(a)), comparable to the out-of-plane structural coherence length of ~29 nm (Fig. S1(c)). Figure 3(a) shows the azimuthal angle (Ψ) dependence of the (1 0 24) magnetic Bragg peak (see Fig. S4(a)) in which the sample is rotated about the scattering vector with Ψ = 0 defined such that the crystallographic $a$-axis lies in the scattering plane as shown in Fig. S3(b). We used the σ - π polarization channel of the incoming and outgoing X-rays because the magnetic Bragg peak intensities follow the relation $I \propto |\vec{m} \cdot (\sigma_i \times \sigma_o')|^2$, where $\sigma_i$ and $\sigma_o'$ are the incoming and outgoing polarizations of photons and $\vec{m}$ is the $J_{\text{eff}} = 1/2$ pseudo-spin magnetic moment [24,25]. We observed that the azimuthal angle dependence of the integrated peak intensity, obtained by fitting the Gaussian peak shape of L-scans, follows the theoretical calculation for $\vec{m} \parallel \vec{b}$. This observation indicates that the magnetic moment is along the crystallographic $b$-axis without any secondary magnetic domains of $\vec{m} \parallel \vec{a}$. Note that this observation of $\vec{m} \parallel \vec{b}$ is consistent with the result of Ref. [18], in which the orthorhombic distortion of ~0.03% is applied using uniaxial compressive strain along the $a$-axis of a $Sr_2IrO_4$ single crystal. These two results confirm that the uniaxial anisotropy due to orthorhombic distortion is crucial in determining the direction of the magnetic moment with respect to the crystallographic axis [26].

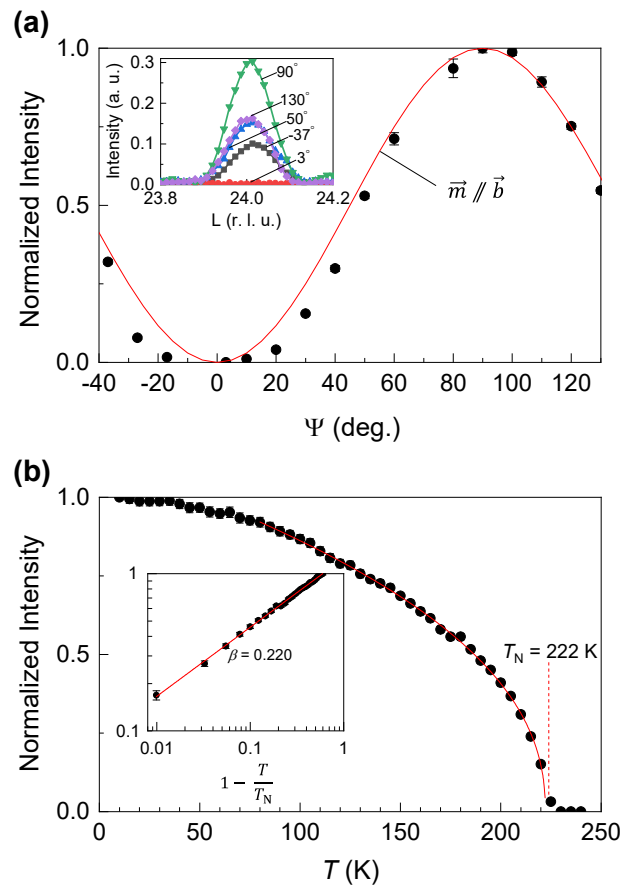

**Fig 3. (a)** Normalized integrated intensities of the Ir L-edge (1 0 24) magnetic Bragg peaks of a Sr$_2$IrO$_4$/Ca$_3$Ru$_2$O$_7$ heterostructure as a function of azimuthal angle (Ψ) with σ - π polarization channel. The solid (red) line represents a theoretical calculation for the $J_{eff}$ = ½ pseudospin magnetic moment ($\vec{m}$) parallel to the crystallographic $b$-axis, as shown in the schematic diagram. Ψ = 0 is defined as the crystallographic $a$-axis lies in the scattering plane. The inset shows the same magnetic Bragg peak intensities at selected azimuthal angles. **(b)** Normalized integrated intensities of the (1 0 24) magnetic Bragg peaks as a function of temperature. The solid (red) line is power law fit using equation $I \propto [1-\frac{T}{T_N}]^{2\beta}$, where $T_N$ and $\beta$ are the Néel temperature (~222 K) and the critical exponent, respectively. (Inset) Normalized integrated intensity vs. the reduced temperature $(1-\frac{T}{T_N})$ in the logarithmic scale. The estimated critical exponent $\beta$ = 0.22 is below 0.23 that corresponds to the 2D XY$h_4$ universality model [30, 31].

Figure 3(b) shows that the integrated intensity of the magnetic Bragg peak (see Fig. S4(b)) decreases with increasing temperature, suggesting the long-range AFM order below $T_N$ = 222 K. Note that this value of $T_N$ is comparable to that of $Sr_2IrO_4$ single crystals (< 230 K) [27-29] despite the orthorhombic distortion. The integrated intensity follows the power-law scaling function $I(t) \approx |t|^{2\beta}$ with the critical exponent $\beta = 0.22$, where $t \equiv 1 - \frac{T}{T_N}$ is the reduced temperature, as shown in the inset of Fig. 3(b). The critical exponent (0.22), corresponding to the 2D XY$h_4$ universality class, seems reasonable for this quasi-2D system with weak interlayer coupling [30-32].

Full (0 1 L) and (1 0 L) scans over the wide range of L from 15 to 25 reciprocal lattice units show clear (1 0 4$n$) and (0 1 4$n$+2) magnetic Bragg peaks with $n$ = 4, 5, and 6, as shown in Fig. 4(a). Additional strong structural Bragg reflections from the $Ca_3Ru_2O_7$ crystal are indicated by asterisks (∗). Note that these magnetic Bragg peaks, i.e., (1 0 4$n$) and (0 1 4$n$+2) with $\vec{m} \parallel \vec{b}$, are the ones expected in the AFM-2 type stacking order, which is observed in orthorhombic systems such as $La_2CuO_4$. In tetragonal $Sr_2IrO_4$ crystals, the AFM-2 type stacking is energetically $2|\delta_c|$ higher than the AFM-1 type stacking, i.e., (1 0 4$n$+2) and (0 1 4$n$) with $\vec{m} \parallel \vec{b}$.

Our experimentally finding not only confirms that the interlayer AFM order is correlated with crystal symmetry but also leads us to further understand that the competition between $j_c$ and $\delta_c$ in determining the long-range AFM order in the system, as schematically shown in Fig. 4(b):

1) The orthorhombic distortion ($\alpha$) = 0, tetragonal $Sr_2IrO_4$ crystals have isotropic $J_{out}$, i.e., $J_{out}^a = J_{out}^b$, resulting in $j_c$ = 0 for a G-type AFM order. Two equivalent AFM domains with the AFM-1 type stacking order with $\vec{m} \parallel \vec{a}$ and $\vec{m} \parallel \vec{b}$ are stabilized with the energy of $\delta_c < 0$. They correspond to *uudd* and *lrrl* stacking order discussed in Ref. [12].
2) $0 < \alpha < \alpha_c$, where $\alpha_c$ is a critical value, as the tetragonal structure of $Sr_2IrO_4$ becomes an orthorhombic structure, the AFM-1 type order with $\vec{m} \parallel \vec{b}$ is stabilized over the other phase with $\vec{m} \parallel \vec{a}$ due to the pseudospin-lattice interaction within the $IrO_2$ basal planes. This transition was experimentally observed with $\alpha \approx 0.03\%$ [18]. $J_{out}$ is no longer isotropic, i.e., $J_{out}^a > J_{out}^b$, resulting in non-zero $j_c$.

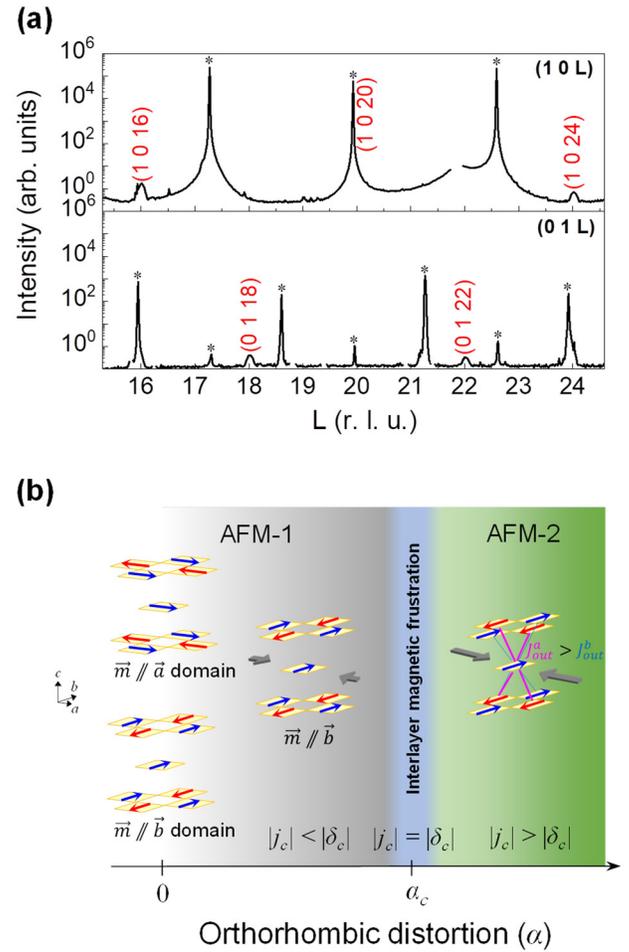

**Fig 4. (a)** Ir L-edge X-ray (1 0 L) and (0 1 L) scans of a $Sr_2IrO_4$/$Ca_3Ru_2O_7$ heterostructure at 6 K. (1 0 16), (1 0 20), (1 0 24), (0 1 18), and (0 1 22) magnetic Bragg peaks are observed. The asterisks [∗] indicate the peaks from the $Ca_3Ru_2O_7$ single crystal substrate. **(b)** Schematic diagram of AFM phases of $Sr_2IrO_4$ as a function of orthorhombic distortion.

Nevertheless, the interlayer pseudo-dipolar interaction remains dominant since $|j_c| < |\delta_c|$.
3) $\alpha > \alpha_c$, the interlayer exchange interaction becomes dominant over the pseudo-dipolar interaction, i.e., $-j_c + \delta_c < 0$, with $|j_c| > |\delta_c|$, making the AFM-2 type stacking order stabilized. Using the tight-binding approximation, we can estimate the value of $j_c$ to be around -0.9 $\mu$eV ($\alpha$ = 1.6%), whose magnitude is larger than the estimated value of $\delta_c$ = 0.3 $\mu$eV.
4) $\alpha \approx \alpha_c$, both AFM-1 and AFM-2 type stacking orders can coexist due to $|j_c| \approx |\delta_c|$, meaning that the system has a large interlayer magnetic frustration and consequently enhances the magnetic quantum fluctuation. By assuming $\delta_c$ = 0.3 $\mu$eV (Ref. [12]) and linear extrapolation between our experimental values, $\alpha_c$ is estimated to be around +0.7%. It would be a formidable task to experimentally study the interlayer magnetic frustration near the critical value. However, applying a gradual strain to obtain continuous orthorhombic distortion in freestanding nanomembranes [33] could provide an accessible approach.

Our result suggests that the competition between interlayer exchange interaction and pseudo-dipolar interaction provides an intriguing tuning parameter to stabilize the long-range magnetic stacking order in quasi-2D systems. We have observed a distinct La$_2$CuO$_4$-like AFM stacking order in Sr$_2$IrO$_4$ by enhancing $|j_c|$ over $|\delta_c|$ via large orthorhombic distortion. In addition to opening exciting questions on the detailed nature of the competition between interlayer interactions, these results raise the prospects of tuning the anisotropic coupling between spins into exotic quantum phases.

**Acknowledgments**

We acknowledge the support of National Science Foundation Grants No. DMR-1454200, DMR-2011876, and DMR-2104296 for sample synthesis and characterization. This research used resources of the Advanced Photon Source, a U.S. Department of Energy (DOE) Office of Science User Facility, operated for the DOE Office of Science by Argonne National Laboratory under Contract No. DE-AC02-06CH11357. Electron microscopy was performed at the Center for Electron Microscopy and Analysis at the Ohio State University. B.K. acknowledges financial support from the Deutsche Forschungsgemeinschaft (DFG, German Research Foundation) – Projektnummer 107745057 - TRR 80.

# Supplemental Materials:

# Emergent interlayer magnetic order via strain-induced orthorhombic distortion in the 5$d$ Mott insulator Sr$_2$IrO$_4$

S. Shrestha[1], M. Krautloher[2], M. Zhu[3], J. Kim[1], J. Hwang[3], J. Kim[4], J.-W. Kim[4], B. Keimer[2], and A. Seo[1]

[1]Department of Physics and Astronomy, University of Kentucky, Lexington, KY 40506, USA
[2]Max-Planck-Institut für Festkörperforschung, D-70569 Stuttgart, Germany
[3]Department of Materials Science and Engineering, The Ohio State University, Columbus, Ohio 43210, USA
[4]Advanced Photon Source, Argonne National Laboratory, Argonne, Illinois 60439, USA

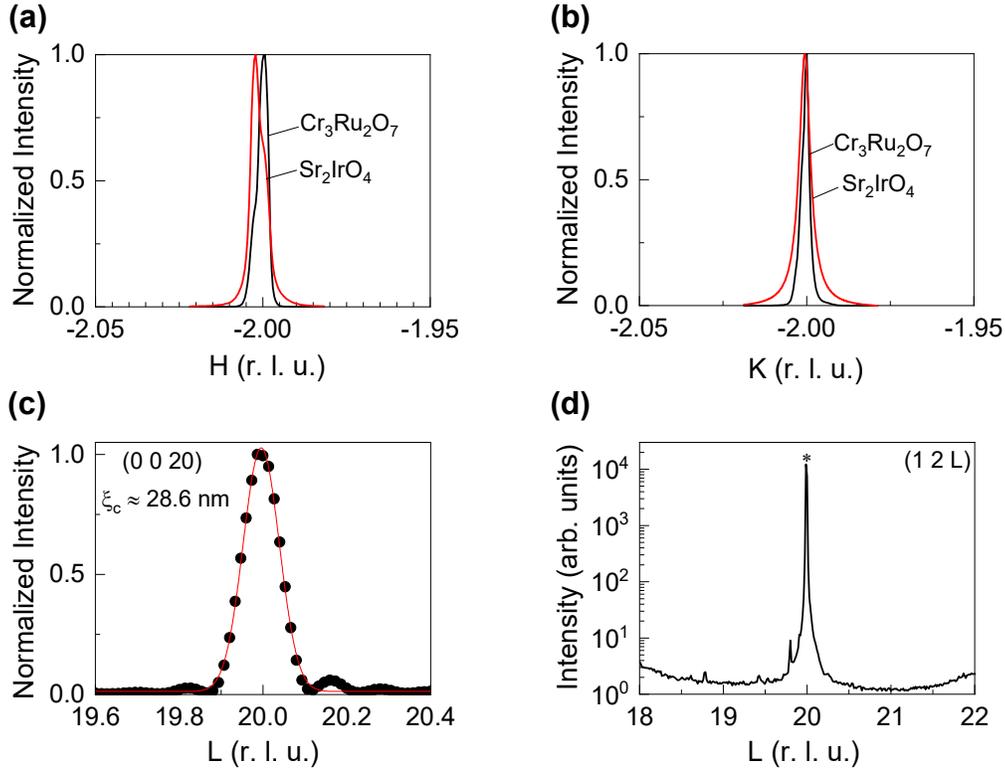

**Fig S1.** **(a)** H scan of Cr$_3$Ru$_2$O$_7$ substrate and Sr$_2$IrO$_4$ thin film measured around the (2 0 16) and (2 0 20) Bragg's reflections, respectively. **(b)** K scan of Cr$_3$Ru$_2$O$_4$ substrate and Sr$_2$IrO$_4$ thin film measured around the (0 2 16) and (and (0 2 20) Bragg's reflections, respectively. **(c)** The L scan of Sr$_2$IrO$_4$ thin film around (0 0 20) structural Bragg's reflection used to measure the coherence length ($\xi_c$) along c-direction. The correlation length was obtained to be 28.6 nm, comparable to the entire 24 nm thickness of Sr$_2$IrO$_4$ thin film. **(d)** (1 2 L) scan of Sr$_2$IrO$_4$ / Ca$_3$Ru$_2$O$_7$ thin film showing no (1 2 odd) peak of Sr$_2$IrO$_4$ thin films. The asterisk [∗] indicates a peak from the Ca$_3$Ru$_2$O$_7$ single crystal substrate.

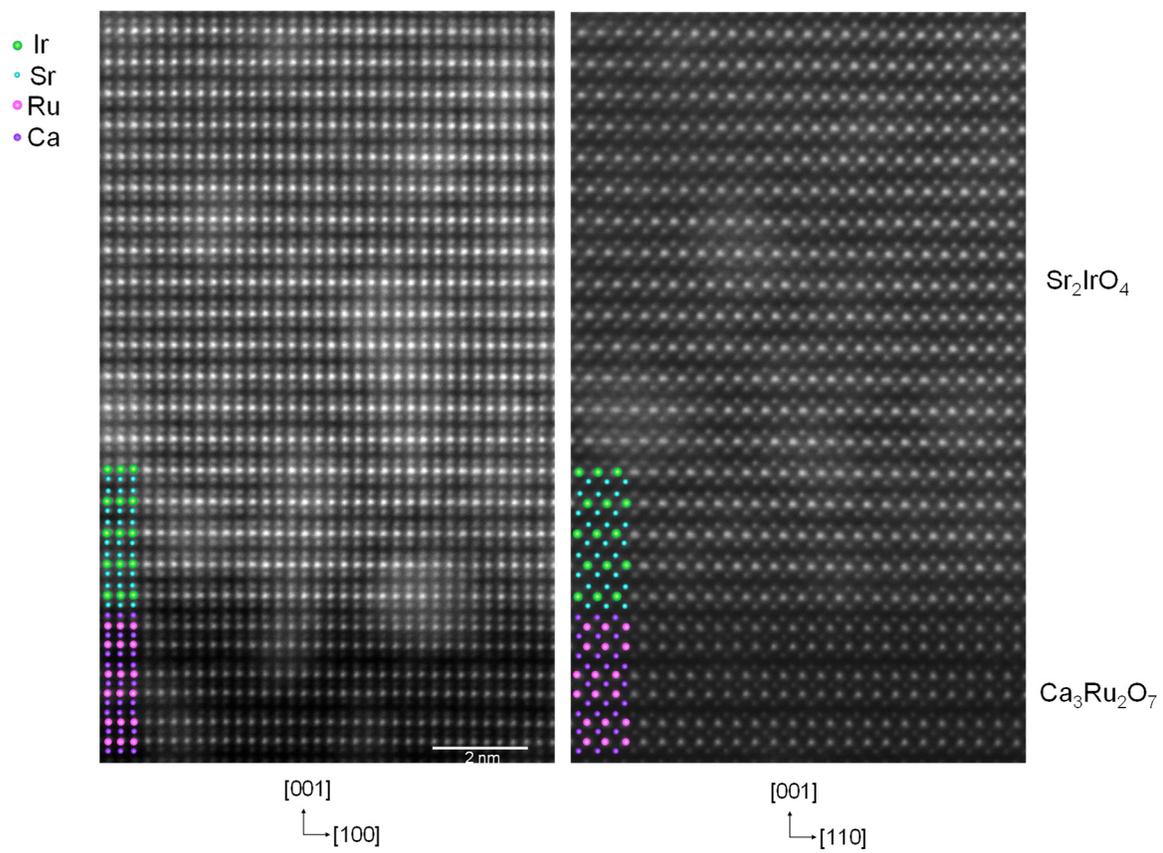

**Fig S2.** High-resolution *Z*-contrast STEM images of a $Sr_2IrO_4/Ca_3Ru_2O_7$ heterostructure along both [1 0 0] and [1 1 0] directions.

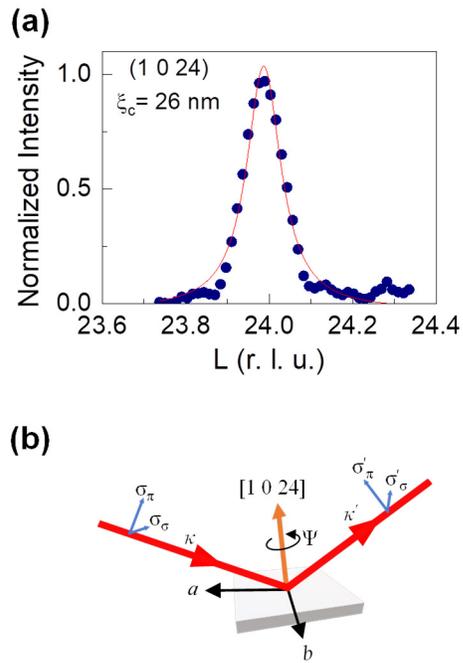

**Fig S3. (a)** The L scan of the (1 0 24) magnetic Bragg reflection. The out-of-plane direction coherence length ($\xi_c$) is obtained to be around 26 nm, comparable to the structural coherence length (Fig. 1S(c)). **(b)** Schematic diagram of scattering geometry. $\sigma$ - $\pi$ polarization channel of the incoming and outgoing X-rays defined by the wave vectors $\kappa$ and $\kappa'$, respectively. $\Psi$ represent the azimuthal angle rotated about the scattering vector where $\Psi = 0$ is defined as the crystallographic *a*-axis lies in the scattering plane.

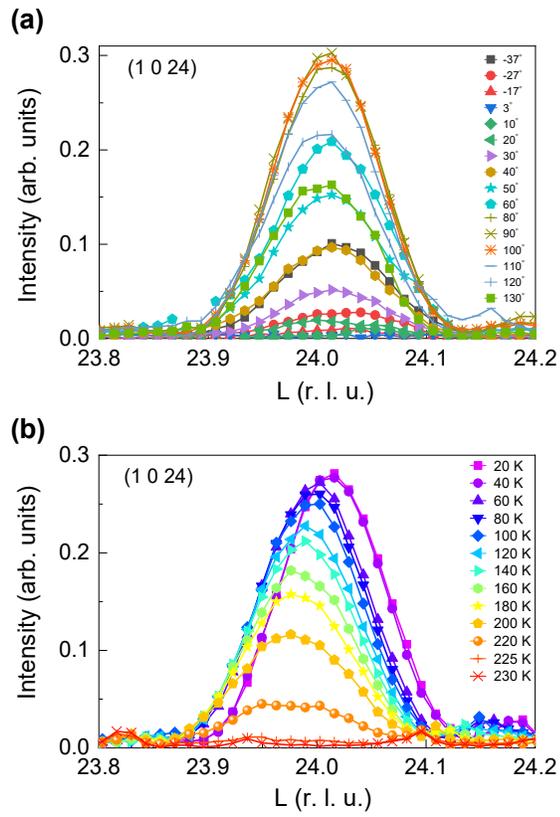

**Fig S4.** The L-scans of the (1 0 24) magnetic Bragg peaks (a) for various azimuthal angles (Ψ) at 6 K and (b) as a function of temperature at Ψ = 90°.